\documentclass[twocolumn,aps,nofootinbib]{revtex4}
\usepackage{amsfonts, amssymb}
\usepackage{graphicx, epsfig,bm}

\newcommand{\be}{\begin{equation}}
\newcommand{\ee}{\end{equation}}
\newcommand{\bea}{\begin{eqnarray}}
\newcommand{\eea}{\end{eqnarray}}

\def\fun#1#2{\lower3.6pt\vbox{\baselineskip0pt\lineskip.9pt
        \ialign{$\mathsurround=0pt#1\hfill##\hfil$\crcr#2\crcr\sim\crcr}}}

\newcommand\lsim{\mathrel{\rlap{\lower4pt\hbox{\hskip1pt$\sim$}}
    \raise1pt\hbox{$<$}}}
\newcommand\gsim{\mathrel{\rlap{\lower4pt\hbox{\hskip1pt$\sim$}}
    \raise1pt\hbox{$>$}}}

\def\dslash{\not{\hbox{\kern-2pt $\partial$}}}
\def\Dslash{\not{\hbox{\kern-4pt $D$}}}
\def\Oslash{\not{\hbox{\kern-4pt $O$}}}
\def\Qslash{\not{\hbox{\kern-4pt $Q$}}}
\def\pslash{\not{\hbox{\kern-2.3pt $p$}}}
\def\kslash{\not{\hbox{\kern-2.3pt $k$}}}
\def\qslash{\not{\hbox{\kern-2.3pt $q$}}}

 \newtoks\slashfraction
 \slashfraction={.13}
 \def\slash#1{\setbox0\hbox{$ #1 $}
 \setbox0\hbox to \the\slashfraction\wd0{\hss \box0}/\box0 }

\def\ee{\end{equation}}
\def\be{\begin{equation}}

\begin{document}
\setlength{\unitlength}{1mm}
%\twocolumn[\hsize\textwidth\columnwidth\hsize\csname@twocolumnfalse\endcsname]
\title{Red Density Perturbations and Inflationary Gravitational Waves}

\author{Luca Pagano,$^1$
Asantha Cooray,$^2$ Alessandro Melchiorri,$^1$ and Marc Kamionkowski$^3$}
\affiliation{$^1$Physics Department and Sezione INFN, University of
Rome ``La Sapienza'', P.le Aldo Moro 2, 00185 Rome, Italy}
\affiliation{$^2$Department of Physics and Astronomy,
University of California, Irvine,  CA 92697}
\affiliation{$^3$California Institute of Technology, Mail Code 130-33,
Pasadena, CA 91125}

\date{\today}%
%\maketitle % use with old revtex !

\begin{abstract}
We study the implications of recent indications for a red
spectrum of primordial density perturbations for the detection
of inflationary gravitational waves (IGWs) with forthcoming
cosmic microwave background experiments.  We find that if
inflation occurs with a single field with an inflaton potential
minimized at $V=0$, then Planck will be able
to detect IGWs at better than 2$\sigma$ confidence level, unless
the inflaton potential is a power law with a very weak power.
The proposed satellite missions of the Cosmic Vision and
Inflation Probe programs will be able to detect IGWs
from all the models we have surveyed at better than
5$\sigma$ confidence level.  We provide an example of what is
required if the IGW background is to remain undetected even by
these latter experiments.
\end{abstract}
\bigskip
%\pacs{PACS Numbers: }

\maketitle

%%%%%%%%%%%%%%%%%%%%%%%%%%%%%%%%%%%%%%%%%%%%%%%%%%%%%%%%%%%%%%%%%%%%%%%%%
%\section{Introduction}
%%%%%%%%%%%%%%%%%%%%%%%%%%%%%%%%%%%%%%%%%%%%%%%%%%%%%%%%%%%%%%%%%%%%%%%%%
%%%%%%%%%%%%%%%%%%%%%%%%%%%%%%%%%%%%%%%%%%%%%%%%%%%%%%%%%%%%%%%%%%%%%%%%%

Inflation's \cite{inflationpapers} predictions of a flat
Universe and a nearly Gaussian and nearly scale-invariant
spectrum of primordial density perturbations
\cite{scaleinvariant} have been confirmed
by a suite of recent cosmic microwave background (CMB)
experiments \cite{cmbexperiments,wmap3cosm,kinney06}. A
third prediction of inflation is a stochastic background of
inflationary gravitational waves (IGWs) \cite{GWs}.  The theory
predicts that the IGW amplitude is proportional to the
cosmological energy density during inflation.  Since this
energy density varies from model to model, the IGW amplitude
cannot be predicted in a model-independent way.  If, however,
IGWs can be detected, they will provide an important
probe of the physics of inflation.

The CMB polarization is perhaps the most promising tool for 
detecting the IGW background. The statistical properties of CMB
linear polarization are fully characterized by two 
sets of spin-2 multipole moments with opposite parities \cite{szeb}. The
magnetic-type modes (B or curl modes) are produced by
IGWs (or ``tensor'' metric perturbations) and
not density perturbations (scalar
metric perturbations), and they do not correlate with
the temperature nor the electric-type-parity modes (E or
grad modes).  A detection of B-mode polarization would thus
provide good evidence for IGWs \cite{marctwo}.

The scalar and tensor perturbations generated by inflation 
have power spectra that are generally well approximated by power
laws: $P_s\left(k\right) \propto k^{n_s}$ and $P_{t}\left(k\right) 
\propto k^{n_{t}}$, respectively, as a function of the spatial
wavenumber $k$.  The amplitudes and spectral indices can be
written in terms of the inflaton potential $V(\phi)$ and its
first and second derivatives $V'(\phi)$ and $V''(\phi)$.  More
precisely, the spectral indices $n_s$ and $n_t$ can be
written in terms of the slow-roll parameters $\epsilon$ and
$\eta$ \cite{marcarthur},
\begin{equation}
     n_s \simeq 1 - 6 \epsilon + 2 \eta, \quad n_t \simeq - 2 \epsilon,
\label{eqn:spectralindices}
\end{equation}
with
\begin{equation}
     \epsilon \equiv {m_{\rm Pl}^2 \over 16 \pi}
     \left({V'\left(\phi\right) \over
     V\left(\phi\right)}\right)^2, \quad
     \eta \equiv {m_{\rm Pl}^2 \over 8 \pi}
     {V''\left(\phi\right) \over V\left(\phi\right)} ,
\label{eqn:eta}
\end{equation}
where $m_{\rm Pl}$ is the reduced Planck mass.  The observed
density-perturbation
amplitude already fixes $V/\epsilon\simeq 6.6\times10^{16}$
GeV.  The tensor-to-scalar ratio $r$ is $r=16\epsilon$.  Since
$n_t$ and $r$ depend only on $\epsilon$, they
satisfy a consistency relation $n_t \simeq - 2 \epsilon =
- r/8$ that can be tested with measurements, if the IGW
background can be detected.

An important new result from recent CMB experiments is 
that a value $n_s=1$ now seems unlikely; the likelihood for its
value peaks closer to $n_s=0.95$ \cite{wmap3cosm,mactavish}.
Although it is still debated whether the data are inconsistent
with $n_s=1$ at the $2\sigma$ level or at the
$3\sigma$ level \cite{kevin}, it is clear that a value $n_s=1$ is
becoming increasingly difficult to reconcile with the data.  
Although a measurement of $n_s$ does not uniquely determine
$\epsilon$, it is natural to expect that $\epsilon$
and $\eta$ are roughly of the same magnitude, or perhaps that
$\epsilon$ is much larger than $\eta$ (if, e.g., the inflaton
potential can be approximated as linear).  Conversely, it would
be unusual to have a value $\eta \gg\epsilon$,
as it would require that the inflaton $\phi$ sit near an
inflection point of $V(\phi)$, an
unusual location.  A value of $\phi$ near an inflection point is
also unworkable in many potentials, as it does not allow the
right number $N_e$ of $e$-foldings between CMB scales and the
end of inflation (which occurs when the potential becomes
sufficiently steep).  If $\epsilon \gtrsim \eta$ (and barring
some unusual cancellation), then a value $n_s \simeq 0.95-0.99$
implies $\epsilon ={\cal O}(0.01)$, which thus implies a
$r\sim0.1$, perhaps within the range of detectability with
next-generation experiments, as argued similarly in
Ref.~\cite{boyle}.

The purpose of this paper is to quantify these arguments 
a bit more precisely by considering a variety of functional
forms for $V(\phi)$, consistent with current data.  As we will
see, the amplitude of the IGW background is large enough to be
detectable by next-generation experiments in a broad family of
inflation models consistent with current data.

\begin{figure}[htbp]
\epsfxsize=3.3in
\centerline{\epsfbox{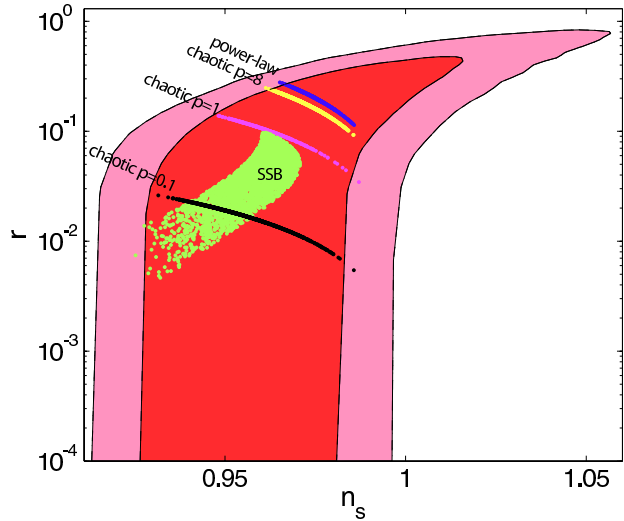}}
\caption{The WMAP (red shaded areas) constraints to the $n_s$-$r$
     parameter space.  The outer (inner) contours are the
     $2\sigma$ ($1\sigma$) contours.  We also plot
     the models within $\Delta 
     \chi^2=1$ ($1\sigma$) from the best fit for power-law and chaotic
     models, and at  $\Delta \chi^2=2.3$ (also $1\sigma$) for the
     SSB models. The curves, from top to bottom, are for
     power-law and for chaotic $p=8$, $p=1$, and $p=0.1$.  The
     (green) points are for the SSB models.}
\label{figure1}
\end{figure}

We consider three classes of inflationary models
\cite{dodelson97,kinney98a,tristan}: (1) {\it
Power-law} inflation is characterized by an inflaton potential
$V(\phi) \propto e^{\phi/\mu}$, where $\mu$ is a mass scale.  In
this potential, there is a relation, $r=8(1-n_s)$, between the
tensor-to-scalar ratio and the scalar spectral index.  (2) {\it
Chaotic} inflation
features an inflaton potential $V(\phi) \propto (\phi/\mu)^p$.
In theoretically attractive models, $p$ is a small integer.
Experimentally, $p\lesssim10$ if $n_s\gtrsim0.9$
\cite{tristan}, but $p$ can empirically be
arbitrarily small.  We consider values of $p$ between $p=0.1$
and $p=8$ in our numerical work.  In these
models, $r=8[p/(p+2)](1-n_s)$.  (3) {\it Spontaneous
symmetry-breaking (SSB)} inflation features an inflaton
potential $V(\phi) \propto [1-(\phi/\nu)^2]^2$.  The precise model is
specified by two parameters: $\nu$ and $N_e$, the number of
$e$-foldings of inflation
between CMB scales and the end of inflation.  A conservative
range for $N_e$ is $47\lesssim N_e\lesssim 62$, corresponding to an
inflationary energy scale in the range of an MeV to $10^{16}$ GeV,
the current upper bound.  The $n_s$-$r$ relation
for SSB models cannot be written in a simple way.  To obtain it,
we use the algorithm given in Eqs. (38)--(41) of
Ref.~\cite{tristan}.  We show in Fig.~\ref{figure1} the
corresponding regions in the $n_s$-$r$ parameter space.

One could also consider hybrid-inflation models, which generally
feature multiple fields.  Their phenomenology can usually be
modeled, though, as a single field with the addition of a
non-zero cosmological constant at the inflaton-potential
minimum.  What distinguishes these models phenomenologically is
that they usually produce $n_s>1$.  Since the working assumption
of this paper is that $n_s<1$, we do not consider these models
further, but simply note that if $n_s$ is indeed greater than
unity, than these models allow for much smaller IGW amplitudes
than the single-field models we consider here.

In their analysis \cite{wmap3cosm}
of their three-year data, the Wilkinson Microwave Anisotropy
Probe (WMAP) collaboration surveyed the cosmological parameter
space assuming $n_s$, $r$, and $n_t$ to be independent
parameters.  The resulting constraints to the $n_s$-$r$
parameter space are shown in Fig.~\ref{figure1}.  However, if
inflation is responsible
for primordial perturbations, and if the
inflaton potential is power-law, chaotic, or SSB, then there is
a fixed relation between $n_s$, $r$, and $n_t$; i.e, they are
not independent parameters.

We therefore re-analyze the CMB data for the power-law,
chaotic, and SSB models, imposing the consistency relations
between $n_s$, $r$, and $n_t$.  To do so, we use the Markov
chain Monte Carlo (MCMC) package \texttt{cosmomcmc}
\cite{Lewis:2002ah} to run a set of chains, imposing these
consistency relations. We use only the latest WMAP results
\cite{wmap3cosm}.  The likelihood is determined using the
October 2006  version of the WMAP likelihood code available at
Ref.~\cite{lambda}. The likelihood are obtained
after marginalizing (with flat priors) over the baryon
and cold-dark-matter densities, the ratio of the 
sound horizon to the angular-diameter distance at decoupling,
and the optical depth to reionization.

In Table~\ref{table2}, we list the current constraints to
$n_s$ and $r$ from WMAP obtained using each class of inflationary models
as a prior.  As expected, the tabulated values and the plots
indicate that tensor modes are required for exponential,
power-law, and small-field scenarios.

\begin{figure}[htbp]
\epsfxsize=3.5in
\centerline{\epsfbox{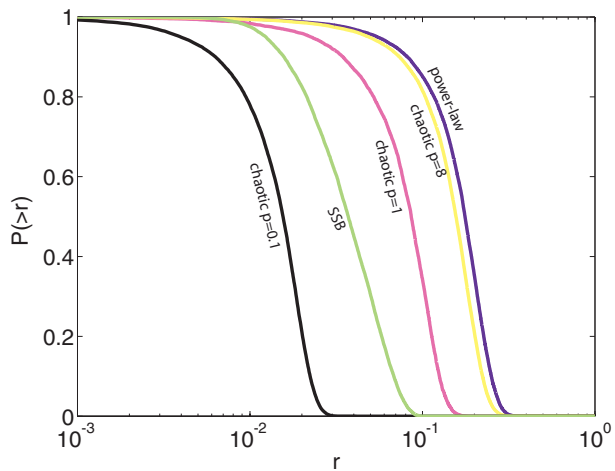}}
\caption{Percentage of inflationary models with a
     tensor-to-scalar ratio above a threshold value $r$, for the
     various inflaton-potential classes considered in
     Fig.~\protect\ref{figure1}. From right to left, the 
     models are power law, chaotic $p=8$, chaotic $p=1$, SSB, and
     chaotic $p=0.1$.}
\label{figure2}
\end{figure}

Our analysis provides a likelihood, from CMB data, for each
point along the curves associated with each class of
inflationary models.
We than thus plot in Fig.~\ref{figure2} the
percentage of models with a tensor amplitude above a threshold
value $r$.  As shown, almost all of the exponential and power-law
models have a tensor amplitude $r > 2\times10^{-2}$, while
small-field models predict $r > 2\times10^{-3}$. 
If experiments were to probe values of $r$ well below $2 \times
10^{-3}$ without finding any evidence for gravitational waves,
then it would rule out a large class of single-field
inflationary models.

\begin{table}[htb]\footnotesize
\begin{center}
\begin{tabular}{r|c|c}
Model & $n_S$ & $r$\\
\hline
Power Law& $0.980^{+0.005}_{-0.005}$ &
$0.16^{+0.04}_{-0.04}$\\
\hline
Chaotic p=1& $0.970^{+0.008}_{-0.008}$ &
$0.080^{+0.020}_{-0.041}$ \\
\hline
Chaotic p=8& $0.978^{+0.011}_{-0.011}$ &
$0.14^{+0.07}_{-0.07}$ \\
\hline
Chaotic p=0.1& $0.964^{+0.019}_{-0.018}$ &
$0.014^{+0.004}_{-0.007}$ \\
\hline
SSB ($N_e=47-62$) & $0.957^{+0.006}_{-0.020}$ &
$0.042^{+0.010}_{-0.014}$ \\
\hline
\end{tabular}
\caption{The $68$\% C.L. limits on the spectral index $n_s$ and
     the scalar-tensor ratio $r$ from WMAP, assuming the
     $n_s$-$r$ relation for each class of inflationary models.}
\label{table2}
\end{center}
\end{table}

Previous work \cite{tensordetect} can already be used to
estimate the fraction of models in each class of models that can
be detected by a given experiment.  Here we go further by
considering values of $n_t\neq0$ that are given by the
inflationary consistency relation.  We also evaluate the
likelihood, given current WMAP constraints to
each class of inflationary models (e.g., power-law, chaotic,
SSB), that IGWs will be detected by a particular future
experiment.

We begin by evaluating, for a given future experiment and for a
given set of inflationary parameters, the corresponding Fisher
Matrix \cite{Jungman:1995av}
\begin{equation}
     F_{ij} = \sum_{l=2}^{l_{\rm max}} \sum_{\alpha,\beta}
     \frac{\partial C_l^{\alpha}}{\partial \theta_i}
     ({\rm Cov}_l)_{\alpha \beta}^{-1}
     \frac{\partial C_l^{\beta}}{\partial \theta_j},
\label{gaussian_fishmat}
\end{equation}
where the $C_l^{\alpha\beta}$ are the power spectra for the
temperature (TT),
temperature-polarization (TE), E-mode polarization (EE), and
B-mode polarization ($\alpha$ and $\beta$ run over TT, EE,
TE, BB), $\theta_i$ is the set of parameters used in the
Markov Chain, and ${\rm Cov}_l$ is the ``spectra covariance
matrix'' \cite{szeb}.  The covariance-matrix entries depend on
the instrumental noise and the angular resolution of the experiment.
The 1$\sigma$ (or 68\% C.L.) error on a particular parameter
${\theta}_j$ is $\sigma(\theta_j)=(F^{-1})_{jj}^{1/2}$
\cite{Jungman:1995av}, after marginalizing over other
undetermined parameters.

Now consider a particular class (e.g., power-law,
chaotic, or SSB) of inflationary models.  The probability that
IGWS will be detected at $N\sigma$, given the current WMAP
constraints, is then,
\begin{equation}
     P=\sum_{i=1}^M \Theta[r_i-N\sigma(r_i)]/M,
\end{equation}
where $\Theta$ is the Heaviside step function.  The sum is over
the $M$ models consistent at $95\%$ C.L. with WMAP assuming
one of class of inflationary models.
For example, if $P=1$ for $N=2$, then the future experiment will
detect IGWs with $2 \sigma$ significance from {\it all} models
compatible with WMAP within that particular class of models.

\begin{table}[htb]\footnotesize
\begin{center}
\begin{tabular}{rcccc}
Experiment & Chan. & FWHM & $\Delta T/T$ & $\Delta P/T$  \\
\hline
WMAP
& 40  & $28'$ & 8.2 & 11.6 \\
$f_{\rm sky}=0.65$
& 60  & $21'$ & 11.0 & 15.6 \\
& 90  & $13'$ & 18.3 & 25.9 \\
\hline
Planck
& 44  & $23'$ & 2.7 & 3.9 \\
$f_{\rm sky}=0.65$
& 70  & $14'$ & 4.7 & 4.7 \\
& 100 & $9.5'$ & 2.5 & 4.0 \\
& 143 & $7.1'$ & 2.2 & 4.2 \\
& 217 & $5.0'$ & 4.8 & 9.8 \\
& 353 & $5.0'$ & 14.7 & 29.8 \\
\hline
BPol
& 45  & $ 900' $ & 0.013 & 0.026\\
$f_{\rm sky}=0.65$
& 70  & $68'$ & 0.12 & 0.24 \\
& 100  & $47'$ & 0.062 & 0.12 \\
& 143  & $47'$ & 0.055 & 0.11 \\
& 217 & $40'$ & 0.046 & 0.092 \\
& 353 & $59'$ & 0.062 & 0.12 \\

\hline
EPIC
& 40& $116'$& 0.032& 0.047 \\
$f_{\rm sky}=0.65$
& 60& $77'$& 0.018 & 0.039\\
& 90& $52'$& 0.0077 & 0.025\\
& 135& $34'$& 0.0073 & 0.036\\
\hline
\end{tabular}
\caption{Experimental specifications for the satellite missions
  considered in this work.  Channel frequency is given in GHz,
  FWHM in arcminutes, and noise in $10^{-6}$.}
\label{table3}
\end{center}
\end{table}

We consider here four experimental configurations: eight years
of WMAP; the Planck satellite \cite{blubuk} (to be launched in
2008); and two proposed next-generation satellites, BPol
\cite{bpol} and EPIC \cite{epic}.  Since the amplitude of the
Galactic foregrounds for polarization is still not determined by
observations, we remove channels below 40 GHz
and above 250 GHz, as these are likely
to be contaminated by synchrotron and dust emission, respectively. 
The parameters assumed for these missions are summarized in
Table~\ref{table3}.

\begin{table}[htb]\footnotesize
\begin{center}
\begin{tabular}{r|cccc}
Model & WMAP8yr & Planck 6ch&BPol&Epic \\
\hline
Power-law& 0.45/0 & 1/0.98 & 1/1& 1/1 \\
Chaotic p=1& 0/0 & 0.99/0.67 & 1/1 & 1/1 \\
Chaotic p=8& 0.30/0 & 1/0.97 & 1/1 & 1/1 \\
Chaotic p=0.1& 0/0 & 0.60/0 & 1/1 & 1/1 \\
SSB ($N_e=47-62$) & 0/0 & 0.78/0.09 & 1/1 & 1/1 \\

\hline
\end{tabular}
\caption{Percentage of models in agreement with the WMAP
     observations and with an IGW background detectable at $2
     \sigma$ and $5 \sigma$ confidence levels by the
     experimental configurations listed in
     Table~\protect\ref{table3}.}
\label{table4}
\end{center}
\end{table}

\begin{table}[htb]\footnotesize
\begin{center}
\begin{tabular}{r|ccc}
Model & 4ch& 100GHzch &allchT+C \\
\hline
Power Law & 1/0.97 & 0.99/0.30 &0.81/0.02\\
Chaotic p=1& 0.99/0.61 & 0.92/0 & 0.25/0 \\
Chaotic p=8& 1/0.96 & 0.99/0.16 & 0.81/0 \\
Chaotic p=0.1& 0.50/0 & 0/0 & 0/0 \\
SSB ($N_e=62-47$)& 0.77/0.07 & 0.36/0 & 0/0 \\

\hline
\end{tabular}
\caption{Planck Only. Percentage of models in agreement with the WMAP
     observations and with a GW-background detectable at $2
     \sigma$/5 $\sigma$ by the Planck experiment in function of
     the channels available for cosmology. The $T+C$ does not
     include the polarization channels but just the temperature
     and the cross temperature-polarization spectrum.}
\label{table5}
\end{center}
\end{table}

Our results are reported in Table \ref{table4}
for detections with $2 \sigma$ and $5 \sigma$ significance.
Our results suggest that
eight years of WMAP data are unlikely to detect IGWs.  The Planck
satellite, on the other hand, stands a good chance
to measure the IGW background in power-law models, even at  the
$5 \sigma$ confidence level. This forecast, however, may be
optimistic, since it assumes perfect
foreground removal and full use of $6$ channels for
cosmology.  To understand these effects better,
we show forecasts in Table~\ref{table5} for Planck,
assuming first a smaller number of channels and then a more pessimistic
configuration where no EE or BB power spectra are used.  While we
found only a small degradation in decreasing the number of
channels from $6$ to $4$, using one channel only, or just
using TT and TE spectra alone, will 
not be able to produce more than an indication for tensor modes. With 
either an Inflation Probe or a Cosmic Vision mission for 
high-precision CMB polarization measurements, we find that the
full set of single-field inflationary models we have considered
can be explored. If no tensor modes are detected after one of
these missions, then it is very likely that inflation cannot be 
described by a single field that rolls to a minimum with zero
cosmological constant.  The alternative would then be more
complicated models, such as chaotic inflation with an unusually
small $p$ ($p\lesssim0.01$), models with multiple fields, string
models, or models with extra dimensions.

%%%%%%%%%%%%%%%%%%%%%%%%%%%%%%%%%%%%%%%%%%%%%%%%%%%%%%%%%%%%%%%%%%%%%%%%%
%%%%%%%%%%%%%%%%%%%%%%%%%%%%%%%%%%%%%%%%%%%%%%%%%%%%%%%%%%%%%%%%%%%%%%%%%
\smallskip
AC acknowledges support from NSF CAREER AST-0645427.
MK acknowledges support from DoE DE-FG03-92-ER40701,
NASA NNG05GF69G, and the Gordon and Betty Moore Foundation.
%%%%%%%%%%%%%%%%%%%%%%%%%%%%%%%%%%%%%%%%%%%%%%%%%%%%%%%%%%%%%%%%%%%%%%%%%
%%%%%%%%%%%%%%%%%%%%%%%%%%%%%%%%%%%%%%%%%%%%%%%%%%%%%%%%%%%%%%%%%%%%%%%%%

\end{document}